# Spatiotemporal Thermal Modulation and Patterning using a Programmable 1024 Element Microheater Array

Rahul Goyal[1, 2], Jang-Hwan Han[1, 2], Sadaf Pashapour[1], and Peer Fischer[1, 2, 3, 4]

[1]Institute for Molecular Systems Engineering and Advanced Materials, Heidelberg University, Heidelberg, Germany
[2]Max Planck Institute for Medical Research, Heidelberg, Germany
[3]Center for Nanomedicine, Institute for Basic Science (IBS), Seoul, Republic of Korea
[4]Department of Nano Biomedical Engineering (NanoBME), Advanced Science Institute, Yonsei University, Seoul, Republic of Korea

Keywords: **microheater, fabrication, electronics, thermal, multiplexer, patterning, printing**

### Abstract

Programmable microheater arrays are essential for a variety of applications including gas sensing, microfluidic lab on a chip devices, 3D printers, and biosensors that rely on DNA amplification. Increasing the density and number of heating elements directly correlates with the precision with which spatiotemporal heat profiles can be delivered. However, large arrays have thus far not been realized. One challenge is that as the number of elements in an array increases, the complexity of connecting them grows. Here, we show that row-column addressing provides a promising architecture for the efficient operation of a large microheater array. We introduce a programmable 32 × 32 microheater array consisting of individually addressable robust platinum (Pt)-based Joule heating elements – each smaller than 300 µm. We show that combining high-voltage multiplexed electronics and sequential addressing controlled by a high frequency clock, allows the independent operation of the 1024 microheater elements. We demonstrate the generation of heat images and the patterning of metallic structures formed from the liquid metal Gallium. Our work demonstrates new capabilities for on-chip thermal devices, and opens the possibility to realize novel heat-controlled micro actuation systems.

## 1 Introduction

Microheater elements with programmable heat generation and spatial control find a number of applications [1, 2], including in haptic displays [3–5], gas sensors [6, 7], actuators and robotics [8–10], 3D printing [11], electronic skin [12], and for encrypted thermal messages [13–15]. The most conventional methodology utilizes pixelated heating elements integrated with direct wire addressing circuits [16]. This approach can in principle generate complex thermal patterns, but has limited scalability. Recently, advanced materials such as multi-layer carbon-fiber-reinforced polymers, carbon nanotubes (CNTs) [17, 18], Fe/Ni/Co alloy (Invar) [19] have also been incorporated in heater designs, thereby broadening potential applications. Still, demonstrations have been limited to directly-wired single heating elements [20] or to small array sizes [21]. The scale-up to larger multi-pixel arrays presents a major challenge, demanding uniform electrical characteristics of the material across all the elements of an array [3, 4, 22], as well as a scalable architecture.

Large numbers of fine, closely spaced heating elements promise the generation of sophisticated heat patterns and potentially steep thermal gradients. However, as the feature size decreases it becomes progressively more difficult to generate sufficient heat, since the resistance drops with shorter conductor lines, and a significant difference in the impedance of the microheater cell and the interconnected input power lines is needed. Moreover, it becomes challenging to address large arrays, as wiring large number of heaters becomes challenging. For instance, wiring each element separately becomes impractical. Herein, we show how high-speed row/column addressing can be used to overcome these challenges and how this enables the selective activation of any individual elements within the array. We thereby demonstrate the fabrication, control and operation of, what to the best of our knowledge is the largest microheater array to date.

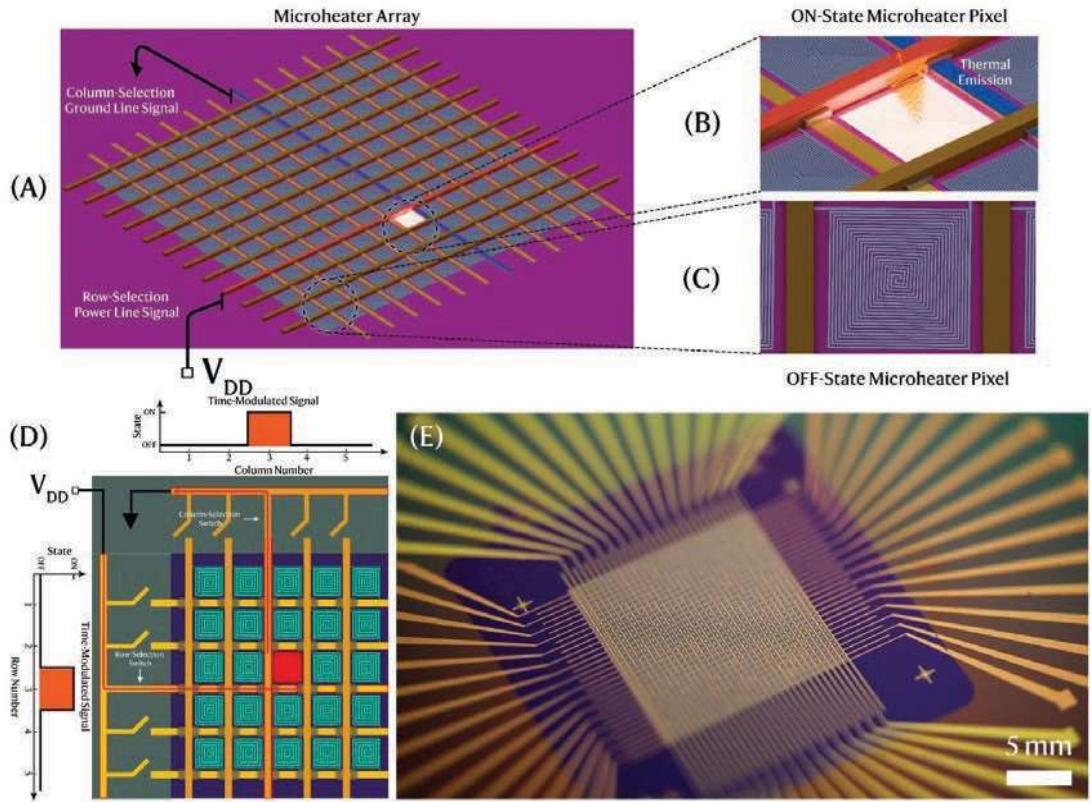

Figure 1: Presentation of fully integrated microheater array for spatial thermal modulation. (A) Schematics of the process of thermal emission in a microheater array with a row-column addressing architecture. (B) The connection of power and ground lines determines the operational state of the microheater pixel, specifically enabling it to function in the (B) ON-State, while ensuring that all other pixels remain in (C) OFF-State. (D) The row- and column-selection switches are externally operated through a time-modulated digital signal that controls the state of pixel. (E) Optical photograph of 2D microheater array fabricated on a three-inch Si/SiO$_2$ (300 nm) wet thermal oxide silicon wafer. The wafer contains a total of 1024 double-spiral square layout-shaped microheaters.

Our microheater array consists of 1024 elements, arranged in a 32×32 closely spaced grid, where each heater has a dimension of 297.5 µm. A schematic with the design and operating concept is illustrated in Fig. 1 (A). Each microheater can be independently addressed to generate spatiotemporal thermal profiles. We adopt a double-spiral square microheater geometry (Figs. 1 (B) and (C)), which provides efficient high-temperature operation and localizes the heat for spatiotemporal thermal control. The modulation is performed through multiplexer electronics. The array has 64 independent control channels, with 32 rows and columns, each equipped with an independent switch control circuit. This permits the direction and polarity of the voltage signal to be controlled, thereby achieving a time-modulated powering of individual microheater elements. Fig. 1 (D) depicts the schematic circuit diagram of a $5 \times 5$ array, consisting of 25 resisting elements for joule heating operations. The entire $32 \times 32$ microheater array is seen in Fig. 1 (E).

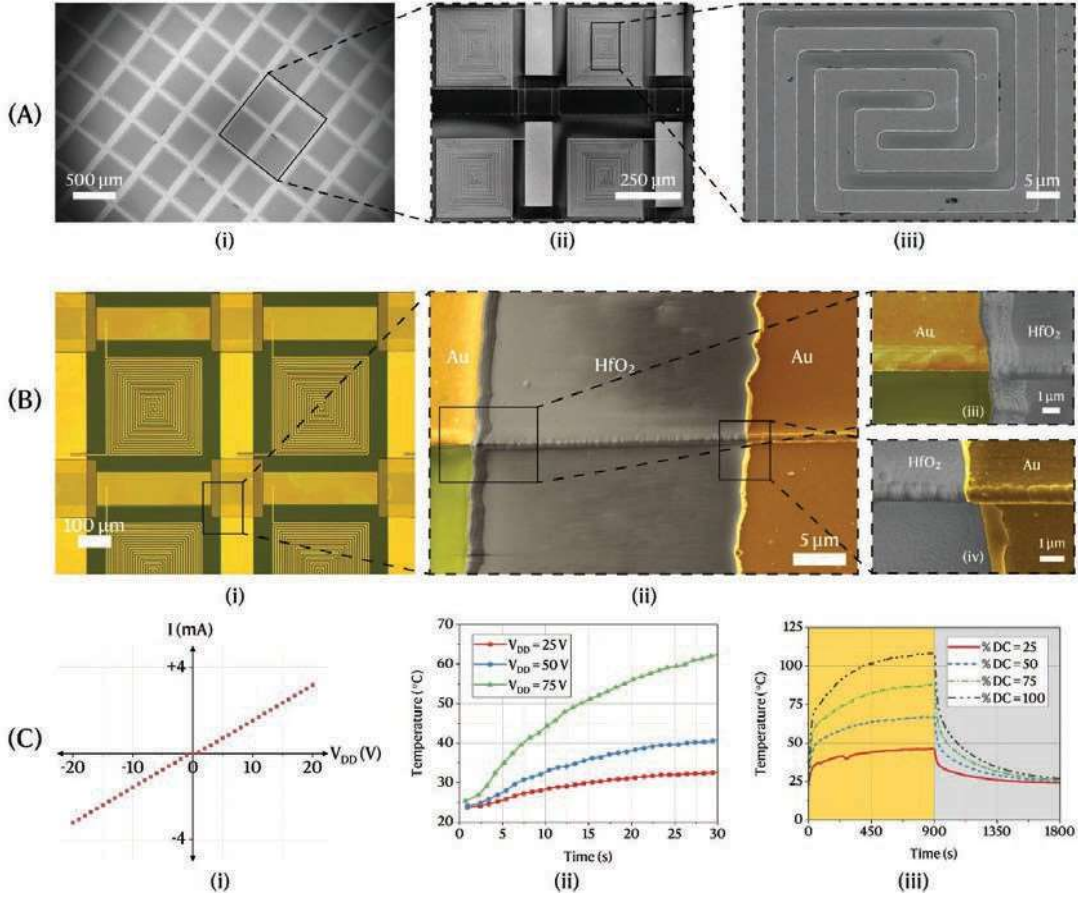

Figure 2: Structure of the microheater array and electrical characterization. (A) Scanning electron microscopy images of the layout of microheater array. (i) Input and output terminals are connected to the square-layout shaped microheaters for multiplexed operation. (ii) Enlarged image of inset, showing a 2 × 2 subset. (iii) Magnified double-spiral structure of a microheater element. (B) False-colored scanning electron microscopy image of the microheater array where $HfO_2$, Au, and $SiO_2$ are shown in grey, yellow and green colors, respectively. (i) Layout of the microheater array connected through row-column architecture. (ii) The overlapped region between input/output bus lines and crossover isolated area of the device. (iii) $HfO_2$ layer isolates the input busline from the output busline. (iv) Output busline patterned over the $HfO_2$ to establish electrical connection with the microheater elements. (C) Experimental measurements for microheater element. (i) Electrical characteristic curve of an element when $V_{DD} \leq 20$ V. (ii) The effect of DC voltage ($V_{DD}$) on the heating performance under 30 s of continuous operation. (iii) The effect of duty cycle (%DC) on the heating and cooling performance of the element under 75 V AC square wave signal.

## 2 Design and Characterization of the Device

An overview of the microheater array and the scheme used to address elements together with images of the device is shown in Fig. 1. Each microheater element in the 32 × 32 square-array has a side length of 297.5 μm and consists of a square-shaped double spiral Pt conductor line. We chose Pt due to its low thermal expansion coefficient (∼ $8.9 \times 10^{-6} K^{-1}$) and high resistivity (∼ $10.1 \times 10^{-8} \Omega$ m). Moreover, Pt provides advantage over other materials which necessitate highly sophisticated fabrication processes that includes electrical interfacing of the advanced materials which substantially complicates the overall system [19, 23–27]. The center-to-center distance between heater elements is 500 μm. The entire array measures 15.5 mm × 15.5 mm, as shown in Fig. 1 (E). Each element has in total 40 variable radius turns with a minimum feature size of approximately 2 μm. The false-color scanning electron microscopy (SEM) images of the device and the corresponding crossover junction are shown in Fig. 2 (B) (i), (ii), and (iii). Since row/column addressing is used (see below for details), the crossing conductor lines are isolated from each other by a 1 μm thick $HfO_2$ layer (see Methods for details). The $HfO_2$ layer can be observed in Fig. 2 (B) (ii), and (iii). The SEM image shows the top Au bus line fabricated over the $HfO_2$ dielectric layer. The experimentally measured current-voltage response of a single microheater element is shown in Fig. 2 (C) (i), and confirms the purely resistive impedance in the operating range (peak-to-peak amplitude of the AC voltage signal is kept constant at max. 120 V). The experimental results of the temperature recorded at different rates of the driving voltage further confirm that the purely resistive heating characteristics are independent of the pulse rate.

Contacting individual heater elements with direct electrical connections is not scalable and next to impossible for a dense microheater array. Instead, we activate individual elements by row/column addressing and use a scheme where the heating elements are supplied with periodic square-wave voltage pulses. This permits the control of the target temperature via the percentage duty cycle (%DC), which corresponds to the fractional ON time ($T_{ON}$) with respect to a maximal signal time of (10 ms), and the total number of pulses applied. The temperature evolution recorded under ambient conditions from a single heating element in response to a square-wave pulsating signal with a constant frequency of 100 Hz and increasing voltages is depicted in Fig. 2 (C) (ii). Next, we demonstrate the temperature profiles when a constant peak voltage of 75 V is applied and the duty cycle is varied. The ON time of 2.5 ms, 5.0 ms, 7.5 ms, and 10 ms, correspond respectively to %DC of 25%, 50%, 75% and 100%. The results are shown in Fig. 2 (C) (iii). The temperature of a heating element will henceforth be controlled by varying the %DC ($T_{ON}$) while keeping the voltage $V_{DD} \leq 75$ V.

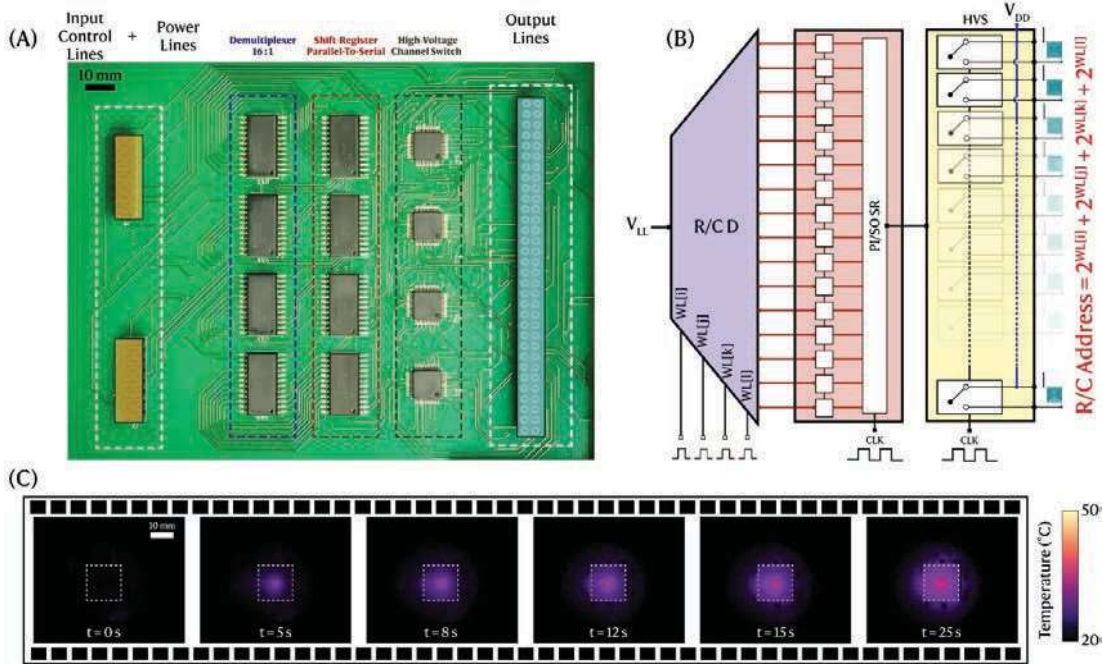

Figure 3: High-speed multiplexing circuit layout used to address individual heater cells in the 32 × 32 array via a serial input driven by an external clock (CLK). In (A) a schematic of the bus-line addressing architecture with the associated bit-transfer sequence is shown. (B) Photograph of the fabricated printed circuit board implementing the corresponding row–column addressing scheme. (C) Infrared images of a single activated heater element at position (16, 16), with the temperature evolution over time as measured in air.

## 3 Multiplexed Operation of Microheater Array

We now describe the row/column addressing architecture to achieve programmable control of the 32 × 32 pixels for a prototype microheater array. Figure 3 (A) shows the image of the printed circuit board that selects the respective row/column of the microheater array for multiplexing. The multiplexer board operates the 32 buslines at 5 bits. Figure 3 (B) shows the signal diagram of the multiplexer board in conjunction with the high voltage switches. More information about the control architecture can be found in the Methods Section 7. Infrared (IR) images captured from the operation of an individual element (16, 16) at different time instants are shown in Fig. 3 (C). The maximum temperature recorded by the IR camera after 25 s of operation is 48.5 °C, where a DC input voltage of 75 V has been provided with a clock frequency of 1 kHz and a pulse repetition rate of 250 pulses/s. The current for a single activated row is approximately 20 mA. When the temperature of a single microheater element in the array is kept at 41 °C, then we find that the heat spreads, but heat spreads, but the full width at half max of the temperature curve has only a corresponding to the width of approximately 4 heating elements"), as shown in Fig. 3 (C). In what follows, multiple heating elements are addressed to form heat patterns.

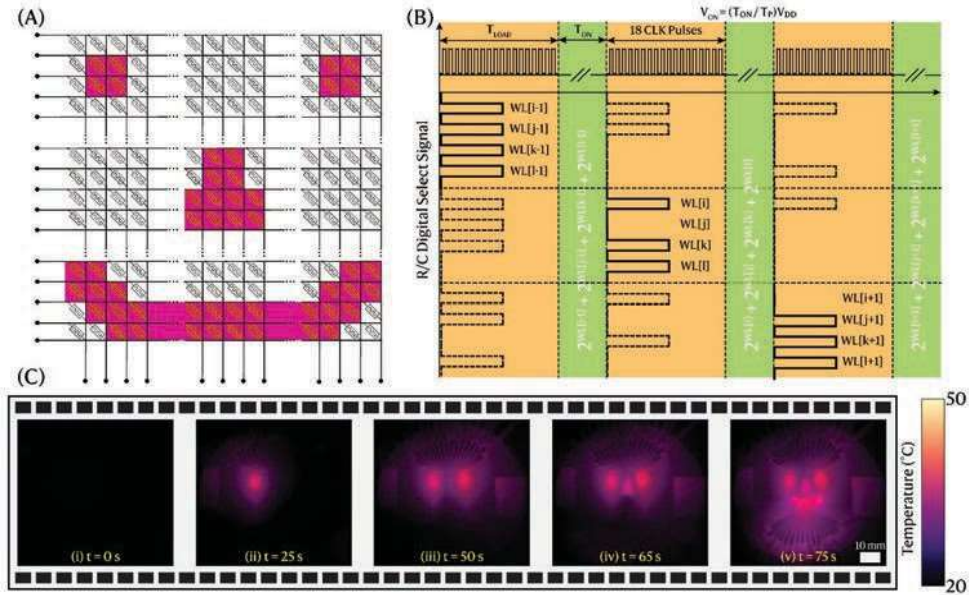

Figure 4: The demonstration of 32 × 32 pixelated microheater array with row/column addressing for complex thermal pattern formation. (A) The circuit model for the pixelated microheater array to generate a complex thermal pattern. (B) The single cell can be activated by switching ON/ OFF the driving voltage line and ground signal line using a R/C digital select signal. The address of the activated pixel element is controlled through 4 bits (WL[i], WL[j], WL[k], and WL[l]) which are loaded in the high-speed multiplexing circuitry using 18 CLK pulses. (C) The infrared images of a complex thermal pattern in the shape of 'smiley face' displayed at various time instants; (i) $t$ = 0 s, (ii) $t$ = 25 s, (iii) $t$ = 50 s, (iv) $t$ = 65 s, and (v) $t$ = 75 s .The minimum and maximum temperatures are 20 °C and 45 °C, respectively.

## 4 Forming Spatial Heat Patterns by Spatially Selective Heating

We now describe how our device is used to form thermal images, as shown schematically in Fig. 4. Specific elements in the array are addressed by loading a four-bit word line which initiates the execution of the corresponding high voltage switch (HVS). For the depicted pattern, in total 18 clock (CLK) pulses, with a CLK signal at 25 kHz and the $T_{ON}$ of 2 ms was used. The times have been estimated based on the number of activated microheater elements and the target temperature. The elements are sequentially addressed using time-division multiplexing as shown in Fig. 4 (B). The control process is schematically shown in Fig. 4 (B). The experimentally recorded sequence of infrared images corresponding to the thermal output of the heater array is depicted in Fig. 4 (C). The relatively low thermal diffusion to neighboring elements resulted in localized heating and a temperature difference between the activated and non-activated microheater elements of at least 15 °C, enabling the formation of distinct thermal patterns.

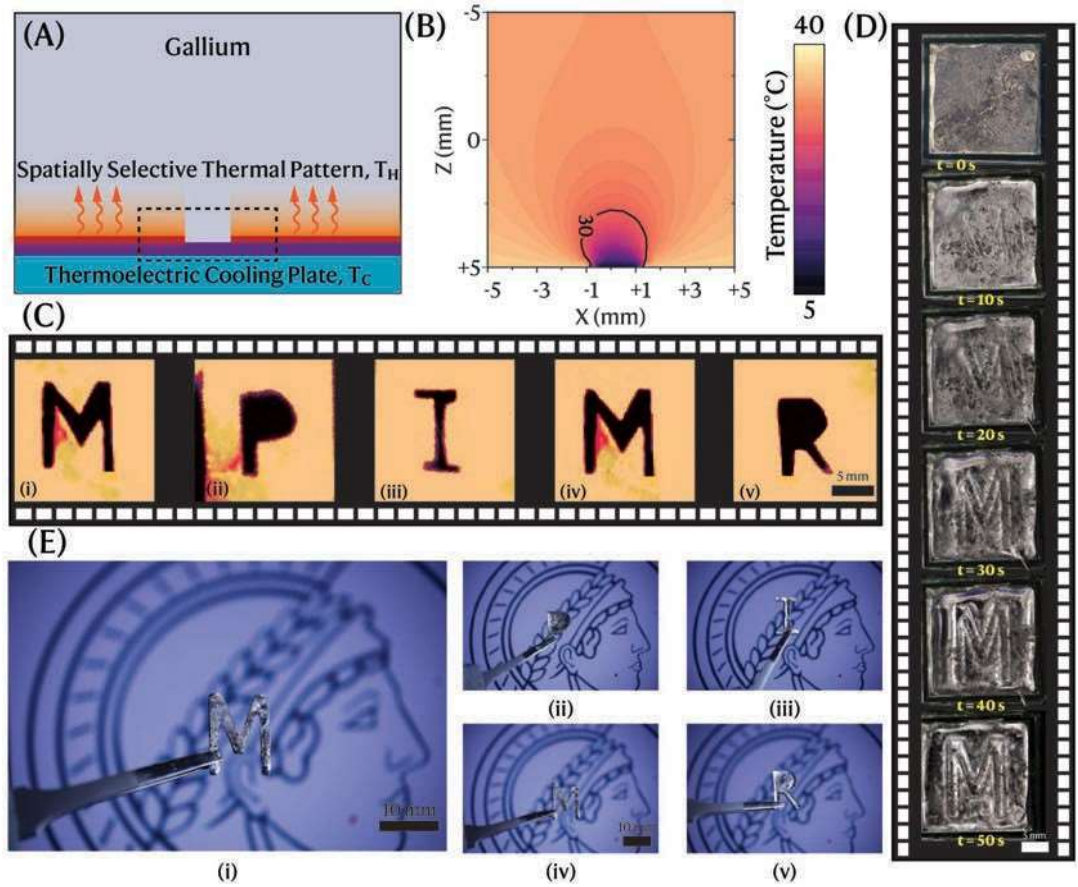

Figure 5: Thermal patterns formed in a liquid metal by the heater array. (A) Schematic illustration of the principle of pattern formation. The thermoelectric cooling plate at temperature $T_C$ is incorporated to ensure a stable base temperature, where the heater elements generate a temperature given by $T_H - T_C$. (B) The numerical results of the heating process simulated with a 2D model showing the surface distribution of the temperature at $t = 50$ s. (C) Sequential infrared images of the thermal patterns resembling the letters: 'M', 'P', 'I', 'M', and 'R'. The results at (i), (ii), (iii), (iv), and (v) show the surface distribution of the temperature for the letters. (D) The optical photographs of 'thermal printing' process at various time points. (E) Photographs of different structures of Gallium-based liquid metal in the shape of the letters (i) 'M', (ii) 'P', (iii) 'I', (iv) 'M', and (v) 'R'.

## 5  Heat Patterning of Structures with Liquid Metals

We demonstrate the operation of the programmable microheater array by forming structures in a liquid metal layer. Precise control over both the temperature and the spatial heat distribution are required to form the letters. Liquid metals are of interest for a number of applications, including flexible displays [28], strain sensors [29], antennae [30, 31], and biomedical sensors [32, 33]. Conventional additive manufacturing enables fast multidimensional fabrication of liquid metal structures with intricate designs for circuitry in flexible electronics [34]. However, liquid metals are challenging to handle due to their material properties. Their high surface tension and the formation of an oxide shell require that printers are customized [35–37] or that the rheological properties of the liquid metal are changed [38–40]. However, these changes may increase the complexity of the additive manufacturing process or deteriorate the electrical properties of the liquid metal. Here, we utilize the relatively sharp solid-liquid transition of Gallium together with the spatiotemporal modulation of the microheater array to pattern the Gallium via a subtractive process. The device is contacted to a thermoelectric cooling plate at a temperature $T_C$ to set a base temperature. The Gallium is solid below 29.76° C. The letters are formed by only heating those elements where the metal is to be removed, which is achieved by melting the Ga and removing the liquid Gallium. The pattering method is schematically shown in Fig. 5 (A). The heat distribution along and extending away from the surface in the area indicated by the black dotted line of Fig. 5 (A) is shown in Fig, 5 (B). The results are shown after 50 s operation of the heater with $T_H$ and $T_C$ corresponding to 35 °C and 5 °C, respectively. We generated Gallium in the shape of the letters 'M', 'P', 'I', 'M', and 'R', as shown in Fig. 5 (C) (i), (ii), (iii), (iv), and (v), respectively. The real-time photographs of the patterning of Gallium captured at different time points are shown in Fig. 5 (D). After the shape is formed, the remaining solid Gallium is detached from the surface using a tweezer. The resultant structures are shown in Fig. 5 (E).

# 6  Conclusions

We have demonstrated the fabrication, operation and control of, what to the best of our knowledge is the largest microheater array to date. Our array consists of 1024 elements, arranged in a 32 × 32 closely spaced grid, where each heater has a dimension of 297.5 µm. Individual elements can be heated up to 150 °C without layer delamination. A double spiral-like heater structure from Pt is used to provide a uniform temperature distribution in each heating element. The wiring challenge is solved by implementing row/column addressing. The control is achieved by a fully programmable high voltage multiplexer board that operates with a maximum refresh time of 3 s which corresponds to the time needed to activate all the 1024 elements of the array and maintain them at 45 °C. The temperature of individual microheater elements is controlled by pulsating the drive voltage and thereby regulating the total current used for Joule heating. Our microheater can be used to generate dynamic heat patterns and we demonstrate the heat-driven melting of Gallium to obtain solid metal shapes. We believe that the scalable architecture demonstrated herein lends itself for applications that require a large number of individually addressable microheaters, as is of interest for applications that benefit from fine spatial temperature control.

# 7  Methods

## 7.1  Numerical Simulation

The spatially selective heating was simulated in MATLAB using an inbuilt finite element model that calculates the variation of heat flux and temperature along the boundaries as a function of time. The values for the thermal conductivity, mass density, and specific heat capacity for Ga are, respectively, as follows: 36.5 W m$^{-1}$ K$^{-1}$, 5910 kg m$^{-3}$, and 371 J kg$^{-1}$ K$^{-1}$. All other material properties were adopted from the built-in library.

## 7.2  Fabrication of the Heater Array

The SiO$_2$/Si substrate was purchased from Siegert Wafer GmbH. The P-doped (Boron) silicon wafer had a 300 nm SiO$_2$ thermal oxide layer. First, the heater elements were fabricated by UV photolithography (Suss, MA6 mask aligner), followed by electron-beam evaporation of Pt and a lift-off process. This was followed by the deposition of the Ti/Au electrode/wire patterns by the same method. The isolation of the conductor lines at the crossover junctions involved the steps of UV photolithography, followed by the electron-beam deposition of 1 µm HfO$_2$. The Ti/Au layers were sputtered sequentially by an electron-beam evaporator (Thermionics). Finaly, the device was coated with a parylene layer (PDS 2010, Specialty Coating Systems GmbH) via chemical vapor deposition using a two temperature step program (∼ 250 °C and ∼ 600 °C). The layer thickness was determined by the amount of loaded parylene dimers (precursor) and was estimated to be around 1 µm. The Ti, Au, and HfO$_2$ material for electron-beam evaporator system were purchased from Kurt J. Lesker GmbH.

## 7.3  Architecture of the Microheater Array

The row and column electrodes of 32×32 heater array are connected to a custom-built digitally controlled electronics board to enable individual addressing of the microheater elements. The primary function of the board is to control the high voltage ($V_{DD}$) channel switch with a low voltage digital signal ($V_{LL}$). The high-voltage switch opens or closes the circuit to generate the target pattern. The switch is normally open and requires a 5-bit code for switching operation which is controlled by a microcontroller (Arduino Mega 2560). The row electrodes are connected to the output terminal of a high voltage amplifier (WMA-300, Falco Systems GmbH) and the column electrodes are connected to ground. The following electronics components are used in the construction of the board: Demultiplexer - CD74HC4067M (Texas Instruments), Shift-Register - SNLS674 (Texas Instruments), and High Voltage Channel Switch - TMUX9616 (Texas Instruments). LPC 6/3-ST-7.62 PCB connectors (Phoenix Contact) are soldered on the board for the DC and HV voltage supply. The thermal pattern is generated by sequentially switching ON the microheater pixels where the 5-bit code is loaded through external CLK pulses. The pixels are heated for a fixed duration based on the temperature requirements of the output. The high voltage signal profile remains unchanged, but the switches are controlled via the 5-bit input code and the CLK pulses.

## 7.4 Digital Signal Processing to Address Heater Elements

Individual heating elements are addressed via the 32 bus lines and an input wordline of total 5 bits. The first bit of the word line represents the chip-select (CS) bit and the remaining bits are WL[i], WL[j], WL[k], and WL[l]. The digital circuit selects the corresponding row/column based on the five input bits, and triggers the high voltage switch (HVS) such that the voltage $V_{DD}$ and ground signals connect to the selected row and column bus lines, respectively. The schematic representation of the electrical architecture is shown in Fig. 3 (A).

The bit-transfer is exercised through external clock (CLK) as shown in the Fig. 3 (A). An image of the device's digital circuit board is shown in Fig. 3 (B), where different integrated components and input/output lines are highlighted with color-dotted lines. The circuit board controls the switching of the $32 \times 32$ microheater array with a five-input logic provided though input control lines. The OFF state of the HVS represents the floating ground signal and the ON state represents $V_{DD}$ and ground signals to R and C respectively. The selected cell will allow current to floe through the microheater element and accordingly generate the thermal response. The serial input to the high voltage channel switch (HVS) connects the voltage signal $V_{DD}$ to the respective buslines (R/C) for the selective operation of microheater pixel. The busline index (R/C) as a function of wordline can be written as $2^{WL[i]} + 2^{WL[j]} + 2^{WL[k]} + 2^{WL[l]}$ where 'ijkl' is the four bit wordline. Multiple microheater elements are addressed sequentially using time-division multiplexing of the clock signal. The array is also equipped with a driving voltage line and a ground line. The selected row and column cells are in accordance with the output addresses sent from the parallel-in/ serial-out shift register (PI/SO SR).

**Supporting Information**

Supporting Information is available from the author.

**Acknowledgements**

This work was supported by the European Research Council under the ERC Advanced Grant Agreement HOLOMAN (No. 788296). The authors thank the Deutsche Forschungsgemeinschaft (DFG, German Research Foundation) under Germany's Excellence Strategy via the Excellence Cluster "3D Matter Made to Order", EXC-2082/1-390761711. The authors thank the Microfabrication and Microfluidics Core Facility (µFluCF) at the Institute of Molecular Systems Engineering and Advanced Materials (IMSEAM) funded partly by the Health + Life Science Alliance Heidelberg Mannheim. The Health + Life Science Alliance provided state funds approved by the State Parliament of Baden-Württemberg.

**Conflict of Interest**

The authors declare no conflict of interest.

**Data Availability Statement**

The data that support the findings of this study are available from the corresponding author upon reasonable request.